\documentclass[aps,pra,reprint,showpacs,superscriptaddress,longbibliography]{revtex4-1}
\usepackage{mathtools,amssymb,graphicx,units}
\usepackage[plainpages=false,pdfpagelabels,colorlinks=true,linkcolor=red,urlcolor=blue,citecolor=blue,pdftitle={Title},pdfauthor={},pdfdisplaydoctitle=true,pdfduplex=DuplexFlipLongEdge]{hyperref}
\usepackage{soul} 
\usepackage[dvipsnames,usenames]{color}
\usepackage[normalem]{ulem}
\usepackage{graphicx}
\usepackage{xcolor}
\usepackage{bbold} 
\usepackage{float}
\emergencystretch=\maxdimen
\usepackage[colorinlistoftodos,prependcaption,textsize=tiny]{todonotes}
\begin{document}

\title{Learning quantum phase transitions through Topological Data Analysis}

\author{Andrea Tirelli} 
\email{atirelli@sissa.it}
\affiliation{International School for Advanced Studies (SISSA),
Via Bonomea 265, 34136 Trieste, Italy}
\author{Natanael C. Costa}
\affiliation{International School for Advanced Studies (SISSA),
Via Bonomea 265, 34136 Trieste, Italy}
\affiliation{Instituto de F\'\i sica, Universidade Federal do Rio de Janeiro
Cx.P. 68.528, 21941-972 Rio de Janeiro RJ, Brazil}

\begin{abstract}
We implement a computational pipeline based on a recent machine learning technique, namely the Topological Data Analysis (TDA), that has the capability of extracting powerful information-carrying topological features.
We apply such a method to the study quantum phase transitions and, to showcase its validity and potential, we exploit such a method for the investigation of two paramount important quantum systems: the 2D periodic Anderson model and the Hubbard model on the honeycomb lattice, both cases on the half-filling.
To this end, we have performed unbiased auxiliary field quantum Monte Carlo simulations, feeding the TDA with snapshots of the Hubbard-Stratonovich fields through the course of the simulations.
The quantum critical points obtained from TDA agree quantitatively well with the existing literature, therefore suggesting that this technique could be used to investigate quantum systems where the analysis of the phase transitions is still a challenge.
\end{abstract}
\maketitle
\section{Introduction}
\label{sec:intro}

A central subject in Condensed Matter Physics and Statistical Mechanics is the study of phase transitions and critical phenomena~\cite{Stanley99,Herbut07}.
In the last decades, due to the increasing computer power resources, numerical methods have become an indispensable tool for the analysis of classical and quantum interacting systems.
Most of these methods, such as Monte Carlo simulations, are performed at finite size systems, which demand the analysis by scaling theories to avoid misleading finite size effects~\cite{Landau21,Sandvik10,gubernatis16,Becca17}. 
However, depending on the type of systems (classical or quantum), or the geometry/dimensionality, performing a finite size scaling (FSS) analysis may be a challenge -- sometimes an unfeasible task --, due to technical bottlenecks: for instance, as a paradigm, in quantum Monte Carlo simulations the occurrence of the infamous minus-sign problem, i.e.~the occurrence a negative statistical weight, restricts the simulations to small lattice sizes~\cite{Loh90,Troyer05,Santos03}.
Another instance is the analysis of three-dimensional systems, in which an extrapolation to the thermodynamic limit is very demanding, even in absence of the sign problem.

In view of this, it is worth developing techniques that could give hints of the existing phases and their phase transitions at finite small system sizes, but, at the same time, could also provide quantitatively reasonable critical points. With the advent of big data analysis, e.g.~with machine learning techniques, a great expectation is placed to this end. Indeed, over the past few years, there has been an effort to develop and benchmark supervised and unsupervised machine learning techniques\,\cite{Dunjko18,Carleo19a,Carrasquilla20}.
In order to provide their usefulness to examine phase transitions, many techniques, such as principal component analysis (PCA)\,\cite{Wang16,Hu17,Costa17a,Wang17a,Wang18a,Khatami19}, t-SNE\,\cite{Wetzel17,Chng18,Zhang19a}, convolutional neural networks\,\cite{Chng17,Broecker17,Carrasquilla17,Kim18,Zhang18c,Khatami20}, restricted Boltzmann machines\,\cite{Nomura17,Huang17,Melko19}, intrinsic dimension analysis\,\cite{Mendes-Santos21a, Mendes-Santos21b}, and persistence homology\,\cite{donato2016,tran2020,Feng20,Olsthoorn20,Leykam21,Kashiwa21}, have been used to identify the critical points in effective (spin and/or fermionic) models.
In summary, their results are promising, despite also suffering with finite size effects.
 
At this point, we recall that since ordered and non-ordered phases exhibit different scaling properties (due to long-range or short-range correlations), this may affect the topology of the (correlation-function) space generated the degrees of freedom of the system.
Hints of this phenomenon are already given by dimensionality reduction techniques, such as PCA.
The investigation of different models have shown that the numbers and the geometries of the projection points clouds in the reduced subspace generated by the leading principal components change drastically around the critical point.
Given these stimulating results, here we contribute to this discussion by developing a pipeline based on Topological Data Analysis (TDA).
This is an unsupervised methodology that combines clustering methods and persistence homology, therefore being able to identify different topological aspects based on the analysis of a space generated (with a given metric) by the data provided as input to the method.

Here we use TDA to examine the snapshots of quantum Monte Carlo simulations, namely via the finite temperature determinant quantum Monte Carlo (DQMC) method. To benchmark this machine learning technique, we investigate the ground state properties of two well-known Hamiltonians: the periodic Anderson model on the square lattice, and the Hubbard model on the honeycomb one, both at the half-filling. The former exhibits a quantum phase transition (QPT) from an antiferromagnetic (AFM) ground state to a singlet (spin-liquid) one, i.e.~an insulator-to-insulator transition. On the other hand, the latter has a metal-to-insulator one, from a semi-metallic to an AFM ground state. Both critical points were extensively investigated in literature by regular FSS analysis, and we consider them as being exact for sake of comparison with TDA ones.

This paper is organized as follow.  The DQMC method is briefly introduced in Section\,\ref{sec:dqmc}, while the details of the TDA is outlined in Section\,\ref{sec:model}.
The results of TDA to the PAM, and the Hubbard model are presented in Section\,\ref{Section:PAM} and \ref{Section:Honeycomb}, respectively.
The Sections are self-contained, with the individual models 
being briefly introduced in each.
Our final remarks are given in Section\,\ref{sec:concl}.

\color{black}


\section{The determinant quantum Monte Carlo method}
\label{sec:dqmc}

First, we recall that for any given Hamiltonian $\widehat{\mathcal{H}} = \widehat{\mathcal{H}}_{0} + \widehat{\mathcal{H}}_{\rm U}$ in which $ [\widehat{\mathcal{H}}_{0},\widehat{\mathcal{H}}_{\rm U}] \neq 0 $, quantum fluctuation effects appear at the ground state and finite temperatures, and these features are described the partition function $\mathcal{Z}= \mathrm{Tr}\,
e^{-\beta\widehat{\mathcal{H}}}$.
For fermionic models, such as those investigated in this work, $\widehat{\mathcal{H}}_{0}$ and $\widehat{\mathcal{H}}_{U}$ correspond to the non-interacting (kinetic) and interacting (Coulomb repulsion) electronic terms, respectively.
However, obtaining the exact partition function is a challenge, due to the non-commutation of the Hamiltonian terms.

The DQMC method\,\cite{Blankenbecler81,Hirsch83,Hirsch85,White89} is an unbiased numerical approach based on an auxiliary-field decomposition of the interaction term, which maps an interacting system onto free fermions moving
in a fluctuating space and imaginary-time dependent potential.
First, by performing a Trotter-Suzuki decomposition, one is able to decouple the
non-commuting parts of the Hamiltonian in the partition function $\mathcal{Z}$.
That is,  $\mathcal{Z}= \mathrm{Tr}\,
e^{-\beta\widehat{\mathcal{H}}}= \mathrm{Tr}\,
[(e^{-\Delta\tau(\widehat{\mathcal{H}}_{0} + \widehat{\mathcal{H}}_{\rm
U})})^{M}]\thickapprox \mathrm{Tr}\,
[e^{-\Delta\tau\widehat{\mathcal{H}}_{0}}e^{-\Delta\tau\widehat{\mathcal{H}}_{\rm
U}}e^{-\Delta\tau\widehat{\mathcal{H}}_{0}}e^{-\Delta\tau\widehat{\mathcal{H}}_{\rm
U}}\cdots]$, where $\beta=M \Delta\tau$, with $\Delta\tau$ being the
grid of the imaginary-time coordinate axis. 
This decomposition leads to an error proportional to $(\Delta\tau)^{2}$, which can be 
systematically reduced as $\Delta\tau \to 0$.
Usually, one should define $\Delta\tau$ small enough in order to these systematic errors become disregardable compared to the statistical ones (from the Monte Carlo sampling).
Proceeding, the two-particle (interacting) terms
$e^{-\Delta\tau\widehat{\mathcal{H}}_{\rm U}}$ are converted in single-particle ones, $e^{-\Delta\tau\widehat{\mathcal{V}}}$, by means of a Hubbard-Stratonovich (HS) transform \cite{Hirsch83}, with the price of adding auxiliary bosonic fields $s(\mathbf{i},l)$ in both real and imaginary-time coordinates, coupled to fermionic degrees of freedom.

After performing the fermionic trace, the partition function becomes
\begin{align}\label{eq:partition_func}
{\cal Z} \sim \sum_{\{s(\mathbf{i}, l)\}}
\, \prod_\sigma {\rm det} 
\big[ \,I + B_\sigma(M) \cdots B_\sigma(1) \, \big]~,
\end{align}
with $I$ being the identity matrix, while $B_\sigma(l) = e^{-\Delta\tau H_{0}} e^{-\Delta\tau V(l)}$.
Here, $H_{0}$ and $V(l)$ are matrix representations of their corresponding fermionic operators.
The final bosonic trace, $\sum_{\{s(\mathbf{i}, l)\}}$, is performed by standard Monte Carlo techniques, with the product of determinants in Eq.\,\eqref{eq:partition_func} being the statistical weight.
More details about auxiliary field QMC methods can be found, e.g.\ in
Refs.\,\onlinecite{Santos03,assaad02,gubernatis16,Becca17}.

\section{Topological Data Analysis}
\label{sec:model}
In this section we shall explain how techniques from Topological Data Analysis (TDA) can be used in the systematic study of topological phase transitions. Data-driven applications of topology to Machine Learning problems started with the seminal work of Carlsson, \cite{Carlsson2009} and have, since then, seen tremendous developments and widespread applications (\textit{e.g.}~in Time Series Analysis \cite{umeda2017, gidea2018} and Computer Vision \cite{bernstein2020}). TDA has only very recently been applied in the study of phase transitions, the first works in the field being \cite{donato2016, tran2020}. The approach outlined here closely resembles the so-called Topological Persistence Machine of \cite{tran2020}, with some significant differences coming from the fact that we leverage on the fact that the data fed to the TDA algorithms come from Monte Carlo snapshots.

The key idea of TDA is that data, gathered together as point clouds, should be thought of as finite samples from underlying geometric objects and that from qualitative information about such geometric objects one could obtain knowledge about the data and understand how it is organised on a large scale. Persistent homology, one of the main techniques in TDA, is the mathematical tool that allows one to infer topological information from a sample of a geometric object.
\subsection{Persistence Homology}\label{subsec:pers}
Let $X$ be a point cloud belonging to a given metric space (\textit{i.e.}~a set endowed with a distance function $d$). Starting from $X$, for any given positive number $\varepsilon$, we can construct a covering of $X$, 
\[
C_{\varepsilon}(X) = \bigcup_{p\in X} B(p, \varepsilon),
\]
given by the union over all points $p$ belonging to the point cloud of the balls $B(p, \varepsilon):=\{y\ |\ d(y, x)\leq \varepsilon\}$. First, note that for $\varepsilon\leq \varepsilon'$ one has that $C_{\varepsilon}(X) \subset C_{\varepsilon'}(X)$. Moreover, as $\varepsilon$ varies, the topology of the space $C_{\varepsilon}(X)$ changes: for example, for sufficiently small $\varepsilon$, the number of connected components (which, in topology, is often refereed to as the 0-\textit{th} Betti number) of $C_{\varepsilon}(X)$ coincides with the number of points in $X$, whereas, for large enough $\varepsilon$ such a number is equal to 1 (each $B(p, \varepsilon)$ in $C_{\varepsilon}(X)$ has a non-empty intersection with $B(p', \varepsilon)$ for some $p' \in X$). The 0-$th$ Betti number of a topological space $Y$, denoted with $b_0(Y)$, is just the first of a sequence of topological invariants associated to $Y$, one for each positive integer $i\in \mathbb{Z}$, where $b_i(Y)$ counts the number of \textit{i-th dimensional holes} in $Y$; for example, $b_1(Y)$ gives the number of (non-trivial) closed loops of $Y$, so if $Y$ is a circle, one has that $b_1(Y)=1$.

We are interested in keeping track of how topological invariants (such as connected components, loops and $b_i(C_\varepsilon(X))$) vary as we vary $\varepsilon$: more specifically, to each of such invariants we assign a birth value $b$ and a death value $d$, so that the invariant appears the first time in $C_b(X)$ and disappears in $C_d(X)$. Therefore, to each topological invariant we can associate a pair of positive numbers $(b, d)$, which we refer to as the \textit{persistence pair} of the invariant. More generally, to a set of topological invariants $\mathcal{I}$, we can associate a \textit{persistence diagram} $\mathcal{D}(\mathcal{I})$ by combining together all the persistence pairs arising elements of $\mathcal{I}$: 
\begin{equation}
    \mathcal{D}(\mathcal{I}) = \{(b_i, d_i) \in \mathbb{R}^2\ |\ i \in \mathcal{I} \},
\end{equation}
where $b_i$ and $d_i$ denote, respectively, the birth and the death of invariant $i$, for $i\in \mathcal{I}$. $\mathcal{D}(\mathcal{I})$ will simply be denoted with $\mathcal{D}$ when $\mathcal{I}$ is fixed and clear from the context (which will always be the case in this work, where we compare different point clouds $X_T$ using persistence diagrams, with a fixed choice of the topological invariants set $\mathcal{I}$).

Therefore, fixing a set of topological invariants $\mathcal{I}$, one can build a map
\begin{equation}\label{top_emb}
X \longrightarrow \mathcal{D}_X,    
\end{equation}
by associating a point cloud $X$ to its persistence diagram $\mathcal{D}_X$, and we shall refer to Eq.\,\eqref{top_emb} as a \textit{persistence embedding}. Such a mapping allows one to compare, from a topological point of view, two different point clouds $X_1$ $X_2$: indeed, let $\mathcal{PD}$ the set of all persistence diagrams arising from $\mathcal{I}$ and assume that one can define a distance function $d$ on $\mathcal{PD}$; then, we can define a distance measure between $X_1$ and $X_2$ as follows: 
\begin{equation}\label{dist_point_cloud}
    \tilde{d}(X_1, X_2) = d(D_{X_1}, D_{X_2}).
\end{equation}
It is clear from Eq.\,\eqref{dist_point_cloud}, that the way one compares two different point clouds $X_1$ and $X_2$ crucially depends on the choice of the metric $d$ between persistence diagrams. We mention here three possibilities for such a choice: 
\begin{itemize}
    \item the \textit{p-Wasserstein distance}, \cite{kerber2017}: given $\mathcal{D}_1$ and $\mathcal{D}_2$ persistence diagrams, it is defined as the infimum over all bijections $\gamma: \mathcal{D}_1 \cup \Delta \rightarrow \mathcal{D}_2 \cup \Delta$ of 
\[
\left( \sum_{x \in \mathcal{D}_1\cup \Delta} || x - \gamma(x)||_{\infty}^p \right)^{1/p},
\]
where $||\cdot ||_{\infty}$ is the standard $\infty$-norm on $\mathbb{R}^2$.
    \item the \textit{bottleneck distance}, which can be obtained from the $p$-Wasserstein distance, by taking the limit $p\rightarrow\infty$: it is the infimum over the same set of bijections of the value
\[
\sup_{x \in \mathcal{D}_1\cup \Delta} || x - \gamma(x)||_{\infty}
\]
    \item the \textit{Betti distance}: given a persistence diagram $\mathcal{D}$, one defines its \textit{Betti curve} as the function $\beta_{\mathcal{D}}: \mathbb{R}\rightarrow \mathbb{N}$ whose value at $s\in\mathbb{R}$ is the number, counted with multiplicity, of points $(b_i, d_i)$ in $\mathcal{D}$ such that $b_i\leq s \leq d_i$; the Betti distance between two persistence diagrams is defined as the $L^p$ distance between the Betti curves $\beta_{\mathcal{D}_1}$ and $\beta_{\mathcal{D}_2}$.
\end{itemize} The set $\mathcal{PD}$ with any of the distances above is a metric space.

\subsection{Fuzzy Spectral Clustering}

Given $n$ point clouds, $X_1, \dots, X_n$ (for the purposes of this work, one can assume $X_i$ as a snapshot of spin configurations at a given temperature~\footnote{At this point, we have to mention that, for our particular case, since we are investigating magnetic phase transitions, which are related to equal-time quantities, we fed the TDA algorithm with Hubbard-Stratonovich configurations of a single time slice, instead of the complete set of $s(\mathbf{i},l)$.},
obtained through a Monte Carlo simulation), one can construct a square matrix $M_X$ of dimension $n$, $M_X=(m_{ij})$, by letting 
\begin{equation}\label{eq:dist_matrix}
m_{ij} = \tilde{d}(X_i, X_j),
\end{equation}
and we shall call $M_X$ the \textit{distance matrix} of the set of point clouds $X_i$, for $i=1, \dots, n$. The intuitive idea is that, the more topologically similar are the point clouds $X_i$ and $X_j$ the smaller is the value $\tilde{d}(X_i, X_j)$. Moreover, $M_X$ can be used, by means of \textit{clustering algorithms}, to group together point clouds that share the same topological features and separate those with an inherently different topology. To this end, one associated to $M_X$ a complete labelled graph $G=(V,E)$ as follows: $V=\{1, \dots, n\}$, $E=V\times V$, where the label on the edge $(i, j)$ is $m_{ij}$. In this way, one translates the problem of clustering point clouds based on their distance into that of identifying communities of nodes in a graph based on the edges connecting them.

A standard technique to achieve it is given by a class of methods called \textit{spectral clustering} algorithms, which use information from the eigenvalues (spectrum) of special matrices built from the graph or the data set.
We refer the reader to Refs.\,\onlinecite{von2007, liu2018} for a review on the theory behind the spectral clustering algorithm. Note that spectral clustering takes as an input a \textit{similarity matrix}, $S=(s_{ij})$, where, typically, $s_{ij}\in \left[0,1\right]$, with $s_{ii}=1$, \textit{i.e.} the self-similarity of a point is the maximum value reached by $s_{ij}$. A distance matrix $M$ (\textit{e.g.} generated through the method above), can be transformed into a similarity matrix by applying the following Gaussian kernel transformation component-wise: 
\[
k(x) = e^{-\frac{x^2}{2\sigma^2}},
\]
where $\sigma$ is an hyperparameter governing the spread of the Gaussian distribution. In this work we shall use a relaxed version of spectral clustering, named \textit{fuzzy spectral clustering}: the result of the integration of the standard spectral clustering algorithm with the fuzzy k-means algorithm~\cite{jimenez2008fuzzy, bezdek1984fcm}. 

After applying fuzzy spectral clustering to the kernel of the matrix $M_X$, we obtain a membership degree function
\[
l=(l_0, l_1): \{X_1, \dots, X_n\} \rightarrow [0, 1]^2, 
\]
such that $l_0(X_i)$ is the membership degree to the first cluster of the point cloud $X_i$, $l_1(X_i)$ is the membership degree to the second cluster of the point cloud $X_i$, and $l_0(X_i)+l_1(X_i)=1$ for all $i$. To identify the critical point, we analyze the sequence 
\begin{equation}\label{eq:member_list} 
\bar{l} = (l_0(X_1), \dots, l_0(X_n)). 
\end{equation}
The crucial point is that this sequence can be segmented in smaller subsequences based on the range of variation of the membership degree; in particular, the critical point will be one of the endpoints (based on the position) of the longest quasi-constant subsequence of $\bar{l}$. Therefore, the critical point can be detected by a simple linear time algorithm that scans $\bar{l}$ and identified the longest quasi-constant subsequence. 

\begin{figure}[t]
    \centering
    \includegraphics[scale=0.65]{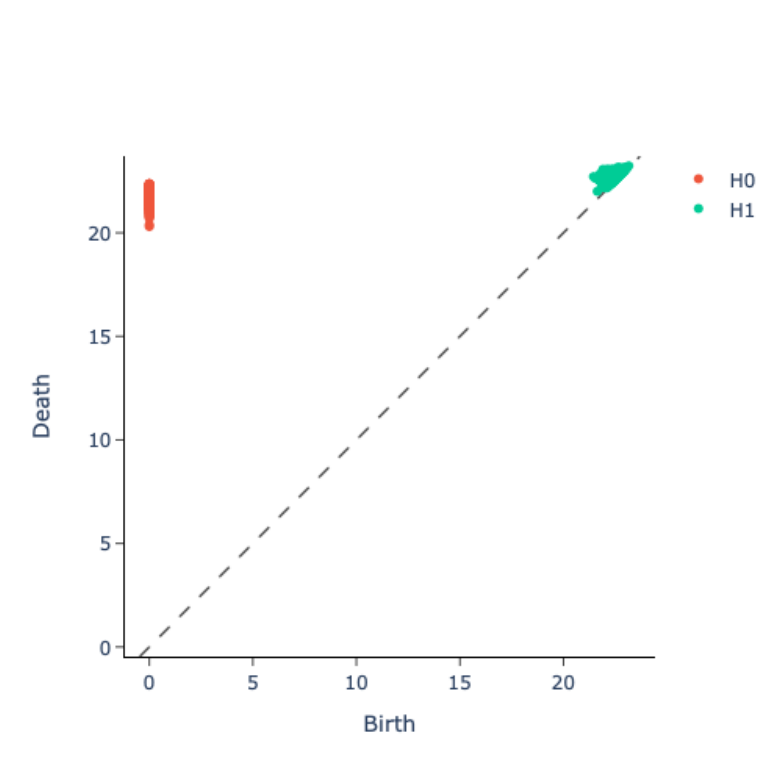}
    \caption{Using the PAM data as point-cloud, for $n=12$ $U=6.0$ and $V/t=1.15$, we show the persistence diagram for the 0-$th$ and first homological dimensions. }
    \label{Fig:pers_diag}
\end{figure}

\subsection{Heuristic Strategy}
The pipeline outlined above clearly depends on a number of hyperparameters, related to the computation of persistence diagrams and the clustering phase. One of the most important and system dependent parameters is the vector of homology dimensions that one decides to compute for the persistence diagram of a certain point cloud $X$. As already mentioned, the $k$-th homological dimensions computes the number of $k$-dimensional \textit{holes} in the manifold underlying to which $X$ belongs. Typically, one starts computing the $0$-dimensional invariants only, as they are the fastest to compute; then, if these do not allow a good separation between the two phases (this is seen in the clustering phase) one starts adding homological dimensions to persistence diagrams. This is generally required for complex systems, where the transition cannot be detected using the simplest topological invariants. 

\section{Results I: the periodic Anderson model}\label{Section:PAM}

We first consider the AF-singlet transition that occurs at the half-filling of the periodic Anderson model (PAM)\,\cite{vekic95,huscroft99,paiva03,hu17b,Costa18b,Schafer19,Zhang19d,Costa19b}.
The PAM is the ``standard model'' to heavy-fermion materials~\cite{coleman07}, in which two species of electrons -- conduction ($d$) and localized ($f$) ones --, experience an adjustable hybridization $V$.
Its Hamiltonian reads
\begin{align}
H = 
-&t \sum_{\langle \mathbf{i},\textbf{j} \rangle \sigma}
\big( d^{\dagger}_{\textbf{i}\sigma}d^{\phantom{\dagger}}_{\textbf{j}\sigma} +
 \mathrm{H.c.} \big)
- V \sum_{\mathbf{i}\sigma} \big( d^{\dagger}_{\textbf{i}\sigma}
f^{\phantom{\dagger}}_{\textbf{i}\sigma}
+ \mathrm{H.c.} \big)
\nonumber
\\
+&U \, \sum_{\mathbf{i}}\big( \, n^f_{\textbf{i}\uparrow}-\frac{1}{2}\,
\big)
\big( \,n^f_{\mathbf{i}\downarrow}-\frac{1}{2}\, \big) \,\, ,
\label{eq:hampam}
\end{align}
where the sums run over a two-dimensional square
lattice with $N=L^2$ sites under periodic boundary conditions.
Here, $\langle \mathbf{i}, \mathbf{j} \rangle$ denotes nearest
neighbors, $t$ is the hopping integral of conduction $d$-electrons, and $U$ is the on-site Coulomb repulsion in the $f$-band, while their
hybridization is $V$.
Within the second quantization formalism, $d^{\dagger}_{{\bf i}\sigma} (d^{\phantom{\dagger}}_{{\bf i}\sigma})$ and $f^{\dagger}_{{\bf
i}\sigma} (f^{\phantom{\dagger}}_{{\bf i}\sigma})$ are standard fermionic creation (annihilation) operators of conduction and localized electrons, respectively, with spin $\sigma$ on a given site $\mathbf{i}$;
$n^f_{\textbf{i}\sigma}\equiv f^{\dagger}_{{\bf
i}\sigma}f_{{\bf i}\sigma}$ are number operators.
We set $t=1$ as the scale of energy.

\begin{figure}[t]
    \centering
    \includegraphics[scale=0.62]{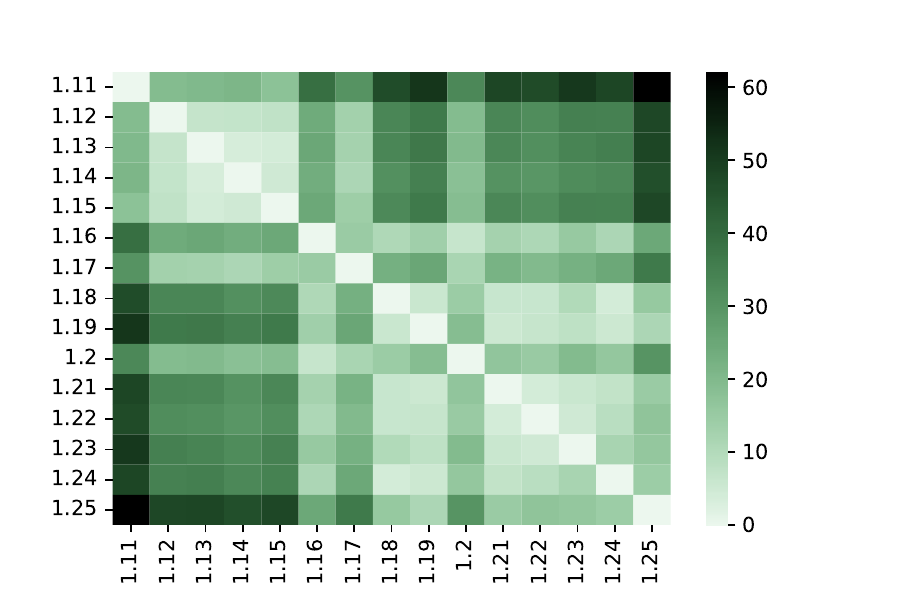}
    \caption{Heatmap representing the distance matrix $M$ for the PAM, for $n=12$ and $U=6.0$, using the Betti distance, with $p=2$; column labels are the different values of the hybridization $V/t$. }
    \label{fig:distance_matrix}
\end{figure}

At low hybridization, the Ruderman-Kittel-Kasuya-Yosida (RKKY) interaction leads to long-range magnetic order, while for large $V$, one observes the formation of local singlets, suppressing the magnetism.
Therefore, a QPT occurs for a critical hybridization $V_c$ separating these
distinct ground states.
Fixing $U/t=4$, FSS analyses provide -- as their best estimate for the critical point, at the present moment -- $V_c/t = 0.99(2)$, and $V_c/t = 1.18(2)$ for $U/t=6$~\cite{hu17b}.
Interestingly, earlier machine learning analysis for the PAM, namely by Principal Component Analysis, provided similar values, but without a reasonable precision.
Given this, throughout this section we benchmark TDA for the quantum phase transition in the PAM.
To carry out the following calculations, needed to implement the methods outlined here, we use the \textit{giotto-tda} \verb|Python| Library\,\cite{tauzin2021}.

We start our analysis by fixing $U/t=6.0$, $\beta t=20$, in a lattice with linear size $L=12$, while varying the hybridization $V/t$.
It is worth to mention that, for this system size, $\beta t=20$ is already a low enough temperature to obtain spin-spin correlations comparable with those of ground state \footnote{Indeed, the correlation length becomes larger than the linear size of the system, therefore, the correlations saturate at lower temperatures}. 
For each value of hybridization, defined for a specified range, $I_V=[V_{lb}, \dots, V_{ub}]$, within which we expect to find the critical $V_c$, we have a point-cloud $X_v$. For this case of $U/t=6.0$, the lower bound of $I_V$ is $V_{lb}=1.11$ and the upper bound $V_{ub}=1.25$, with step size $\Delta V = 0.01$. Here, and in the other following cases, we provide to the TDA algorithm 400 independent Monte Carlo snapshots for each individual choice of parameters ($\beta$, $U$, and/or $V$). Throughout the simulations, these snapshots are obtained to every 100 Monte Carlo sweeps. We have ensured their independence by performing global moves for the Hubbard-Stratonovich fields\,\cite{Scalettar91a}, which drastically reduces the autocorrelation time.

First of all, for each point cloud $X_v$, we compute the corresponding persistence diagram $\mathcal{D}(X_v)$. In this step, the main parameter that has to be chosen is the number of homological dimensions to compute. In \textit{giotto-tda} this is implemented in the class \verb|VietorisRipsPersistence|.
For instance, Fig.\,\ref{Fig:pers_diag} displays an example of such a calculation, for the PAM at $V=1.11$, in which we compute the 0-$th$ and first homological dimensions, depicted with red and green points, respectively.

\begin{figure}[t]
    \centering
    \includegraphics[scale=0.20]{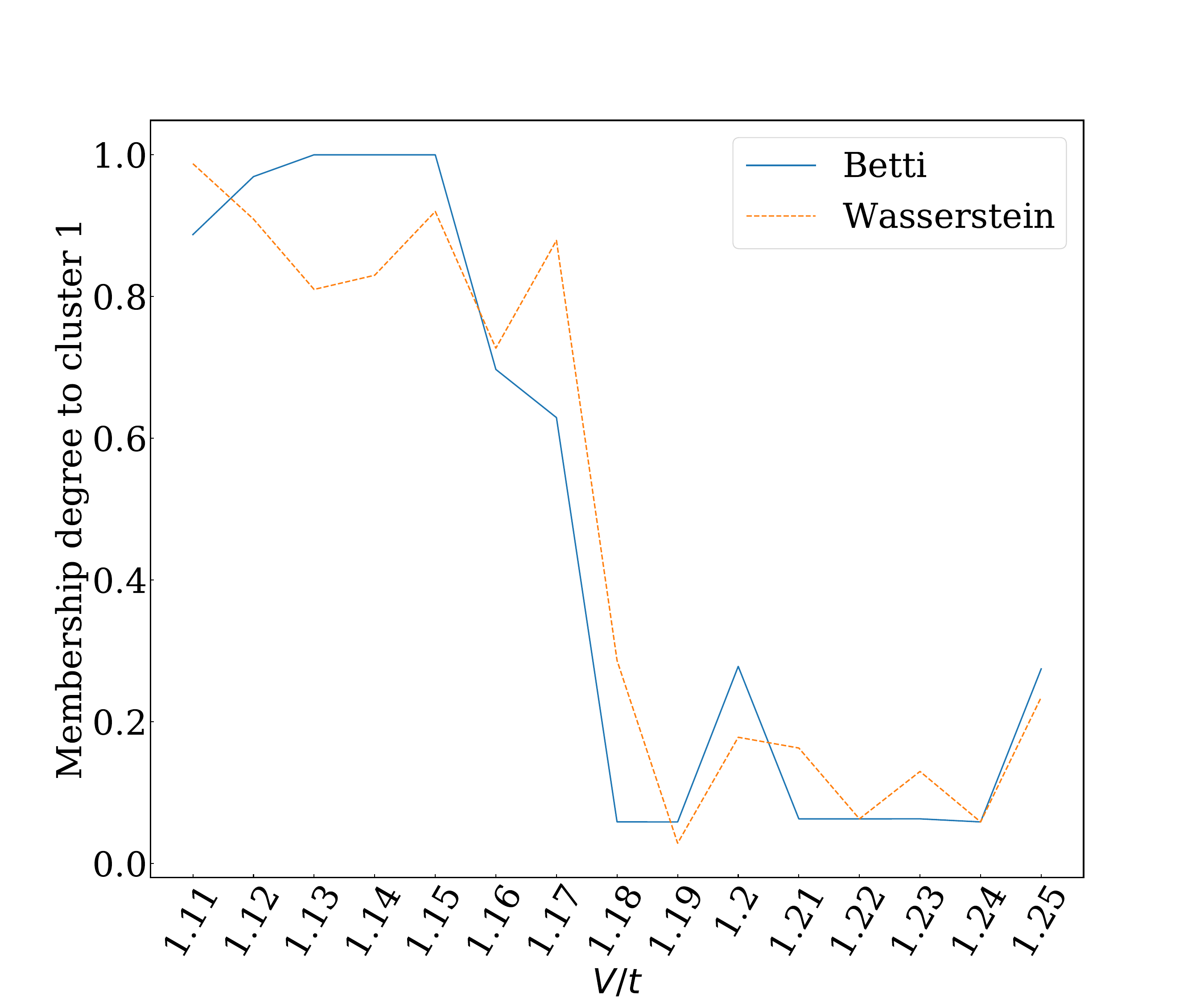}
    \caption{Line-plot representing the membership degree vectors $\bar{l} $to the fuzzy cluster 1 for the PAM, for $n=12$ and $U=6.0$, using the Betti distance (blue solid line) and the Wasserstein distance (orange dashed line), with $p=2$; labels on the $x$-axis are the different values of the hybridization $V/t$.}
    \label{fig:mem_deg}
\end{figure}

Once all the persistence diagrams have been calculated, we have to compute their distance matrix, Eq.\,\eqref{eq:dist_matrix}. As outlined in Subsection \ref{subsec:pers}, there are multiple choice for the distance between two persistence diagrams. In this example, we construct the distance matrix $M$ using the \textit{Betti distance}, with $p=2$. This calculation is implemented in the \textit{giotto-tda} class \verb|PairwiseDistance|.
Figure \ref{fig:distance_matrix} exhibits a heatmap representation of such a distance matrix: as expected, one can visually identify two darker clusters in the bottom-left and upper-right zones, and also two lighter clusters in the upper-left and bottom-right zones. This suggests that: (1) the point-clouds $X_v$ for $V/t$ in the range from $1.11$ to $1.16$ are all topologically similar as well as point clouds in the range from $1.17$ to $1.25$;
(2) a point cloud $X_{v_1}$ for $1.11 \leq V/t \leq 1.16$ is topologically different from all point-clouds in the range [1.17, 1.25];
(3) a point cloud $X_{v_2}$ for $1.17 \leq V/t \leq 1.25$ is topologically different from all point-clouds in the range [1.11, 1.16].

\begin{figure}[t]
    \centering
    \includegraphics[scale=0.31]{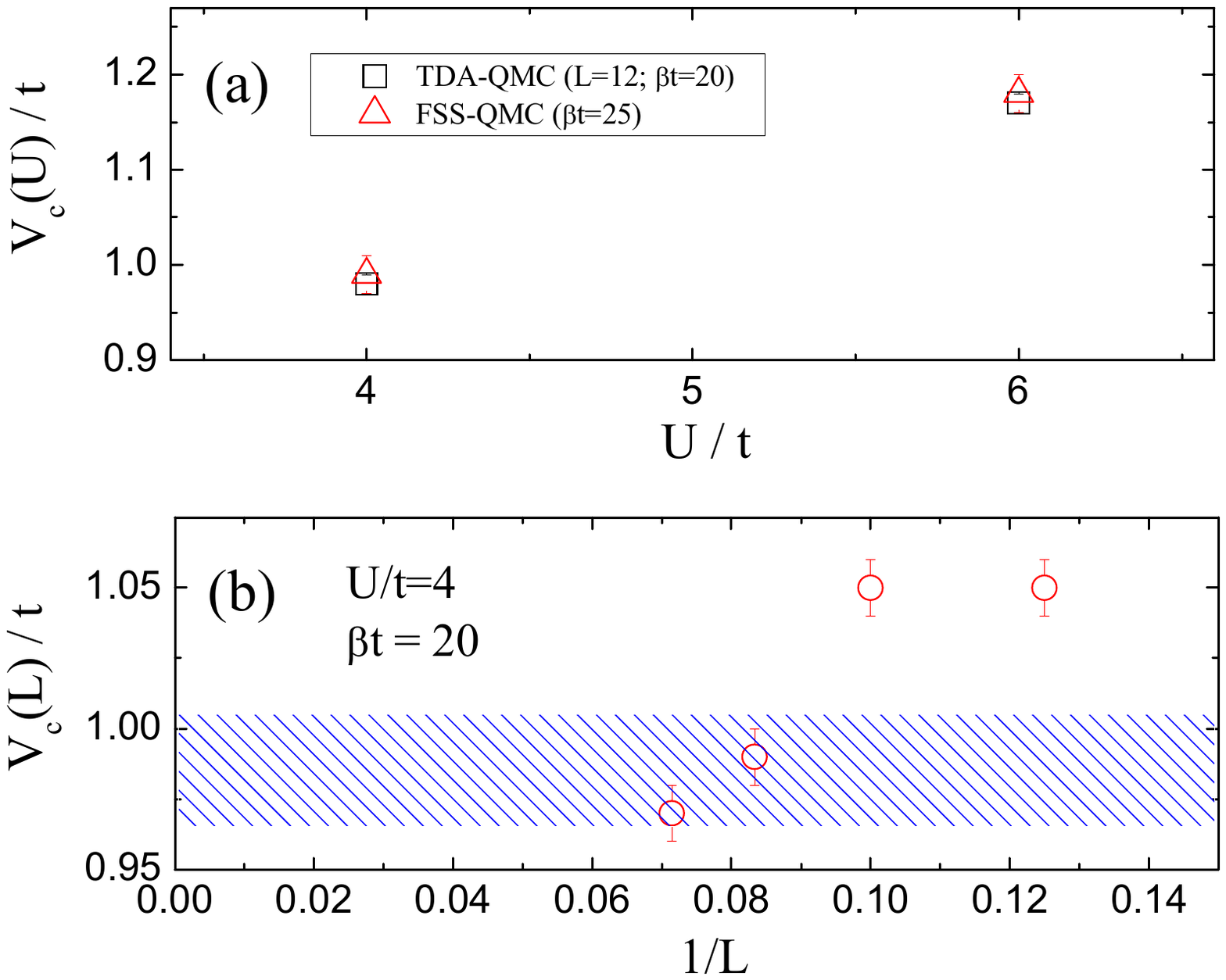}
    \caption{Quantum critical hybridizations for the Periodic Anderson model from Topological Data Analysis. Panel (a) exhibits the TDA $V_{c}(U)/t$ (black square symbols) for fixed lattice size $L=12$, and $\beta t = 20$, comparing them with the results for the FSS analysis in Ref.\,\onlinecite{hu17b} (red triangle symbols). Panel (b) displays the TDA $V_{c}/t$ (red circle symbols) for fixed $U/t=4$, and $\beta t = 20$, and different lattice sizes. The blue hatched area corresponds to the result (within error bars) for the FSS analysis in Ref.\,\onlinecite{hu17b}.}
    \label{Fig:Pam02}
\end{figure}

To formalize the above considerations, we perform a fuzzy spectral clustering analysis on the distance matrix. We implemented it using a combination of functions coming from the \verb|Python| packages \verb|scikit-learn|, \cite{scikit-learn}, and \verb|skfuzzy|, \cite{skfuzzy}. As mentioned earlier, applying fuzzy spectral clustering gives a membership-degree list $\bar{l}$, Eq.\,\eqref{eq:member_list}, which is displayed in Fig.\,\ref{fig:mem_deg}. We have performed the clustering algorithm over two different distance matrices: one is built using the Betti distance, and the other with the Wasserstein distance, both with $p=2$ -- the analysis for $p>2$ does not lead to any significant changes for the construction of the membership lines. Although the fuzzy values in the two cases are not the same, both curves display the same qualitative behavior, i.e.~being drastically reduced around the critical point. In other words, it suggests that the TDA is robust with respect to the choice of distance metric between persistence diagrams. Given this and the fact that implementing Wasserstein distances is more computationally expensive than the Betti one, we use the latter in all the other cases hereafter. Moreover, notice that, almost all membership degree values at $V/t$ lower than $1.17$ are very close to 1 (which is the maximum), whereas, for $V/t \geq 1.17$, most of $\bar{l}$ results are close to zero. A quasi-subsequence detection algorithm detects this behaviour and separate the two zones, giving $V_c=1.17(1)$.
Here, the error bar corresponds to the step size of the hybridization in our analysis, i.e.~$\Delta V=0.01$.
Repeating the same procedure to $U/t=4$, we obtain $V_c=0.98(1)$, as displayed in Fig.\,\ref{Fig:Pam02}.
It is remarkable the agreement between TDA and the standard FSS analysis for the PAM. 

It is also worth to examine the impact of finite size effects to these TDA critical points.
To this end, we have performed the same analysis for system sizes $L=8$, 10, and 14, fixing $\beta t =20$ and $U/t=4$, within the range $V/t=[0.93, 1.07]$.
The critical points $V_{c}(L)$ are displayed in Fig.\,\ref{Fig:Pam02}\,(b), with error bars corresponding to the step size of the hybridization.
As expected, our TDA exhibits finite size effects that could lead to deviations in $V_{c}(L)$, in particular for small systems sizes.
That is, in order to obtain precise results within TDA, the analysis should be done for large $L$.
Even though, this method is promising, since the relative error $\epsilon =  |V_{c {\rm (TDA)}} - V_{c {\rm (FSS)}}|/ V_{c {\rm (FSS)}}$ is small; e.g., for $L=8$ we find $\epsilon = 6(2)\%$.

\section{Results II: the Hubbard model in the honeycomb lattice}
\label{Section:Honeycomb}

Now we proceed investigating the metal-to-insulator QPT in the Hubbard model on the half-filled honeycomb lattice.
The Hubbard Hamiltonian reads
\begin{align}
H = -&t \sum_{<\textbf{i},\textbf{j}>\sigma}
\big( d^{\dagger}_{\textbf{i}\sigma}d^{\phantom{\dagger}}_{\textbf{j}\sigma}+
d^{\dagger}_{\textbf{j}\sigma}d^{\phantom{\dagger}}_{\textbf{i}\sigma}\big)
\nonumber
\\
+&U\sum_{\textbf{i}}\big( \, n^d_{\textbf{i}\uparrow}-\frac{1}{2}\,
\big)
\big( \,n^d_{\textbf{i}\downarrow}-\frac{1}{2}\, \big) ,
\label{eq:hamhub}
\end{align}
in which we adopt the same notation of the previous case.
The Hubbard model in such geometry geometry is of particular interest in condensed matter, since it leads to interacting Dirac fermions.
Many QMC works in literature~\cite{paiva05,Herbut06,Assaad13,Parisen15,otsuka16,Raczkowski20,Costa21} have provided the occurrence of a QPT $U_{c}$, with the emergence of an antiferromagnetic ground state at half-filling.
By contrast with PAM, where both magnetic and singlet phases are insulators, here we have a metal-to-insulator transition.

\begin{figure}[t]
    \centering
    \includegraphics[scale=0.28]{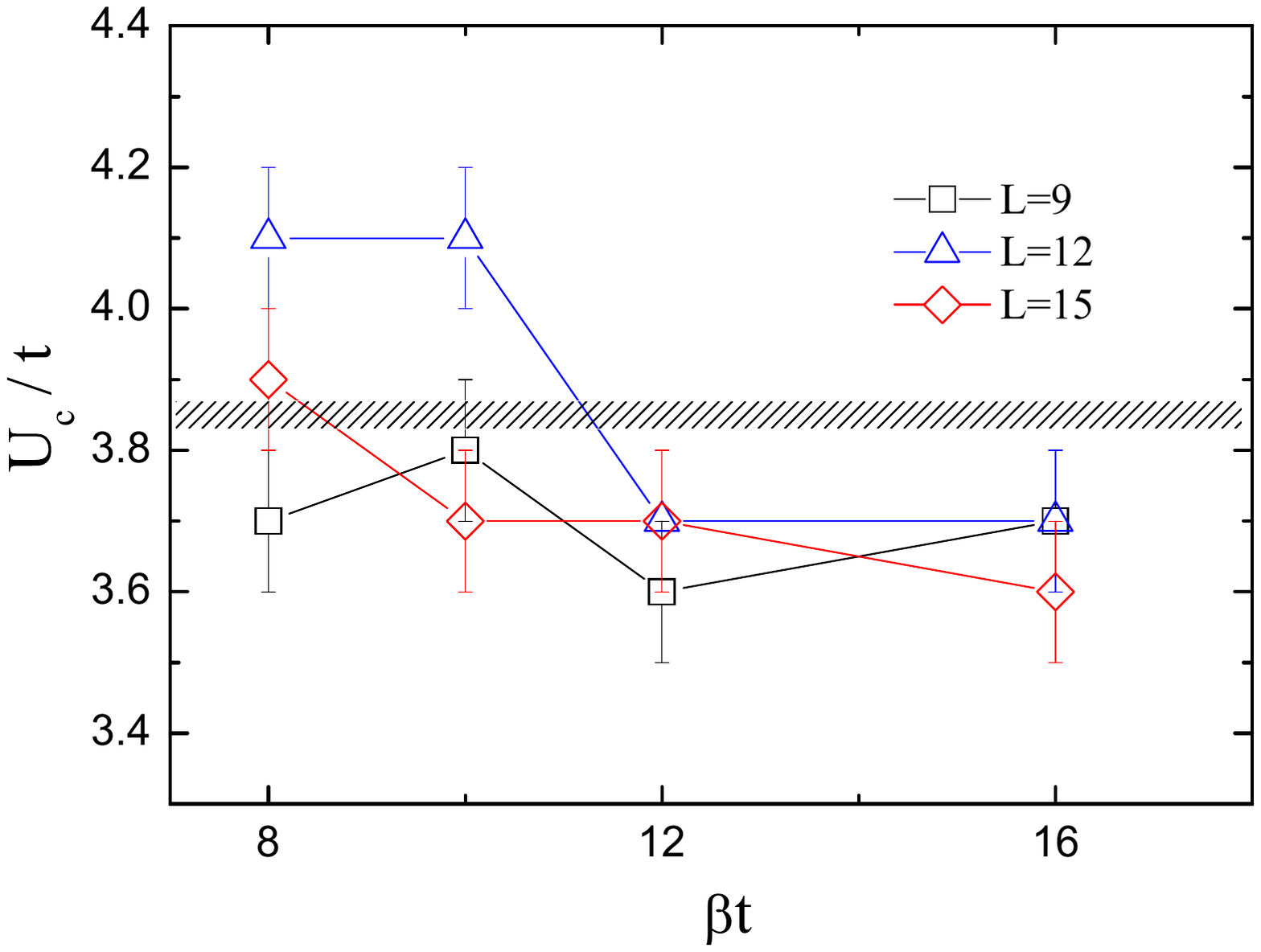}
    \caption{Quantum critical points of the Hubbard model on the honeycomb lattice from Topological Data Analysis as function of the inversion of temperature $\beta$, and for different lattice sizes. The black hatched area corresponds to the result (within error bars) for the FSS analysis in Ref.\,\onlinecite{otsuka16}.}
    \label{Fig:Honeycomb}
\end{figure}

At the present moment, the best estimation to the critical point for this model is $U_c \approx 3.85(2)$\,\cite{otsuka16}, obtained by a FSS analysis for lattice sizes up to $L=36$.
Here, we use TDA to investigate such a QPT for systems sizes $L=9$, 12 and 15, while varying $U/t$ within the range $[3.0, 4.6]$, for a step size $\Delta U/t = 0.1$.
Repeating the procedure outlined in the previous sections we obtain $U_{c}(L,\beta)$, as displayed in Fig.\,\ref{Fig:Honeycomb}.
First, notice that, when temperature is reduced ($\beta$ is increased), the TDA critical points stabilize with a region around $U_{c}/t \approx 3.7(1)$.
Second, different from the PAM, here the TDA critical points are less dependent on the system size, which could be an artifact of the larger step size we chose.

\begin{figure}[t]
    \centering
    \includegraphics[scale=0.60]{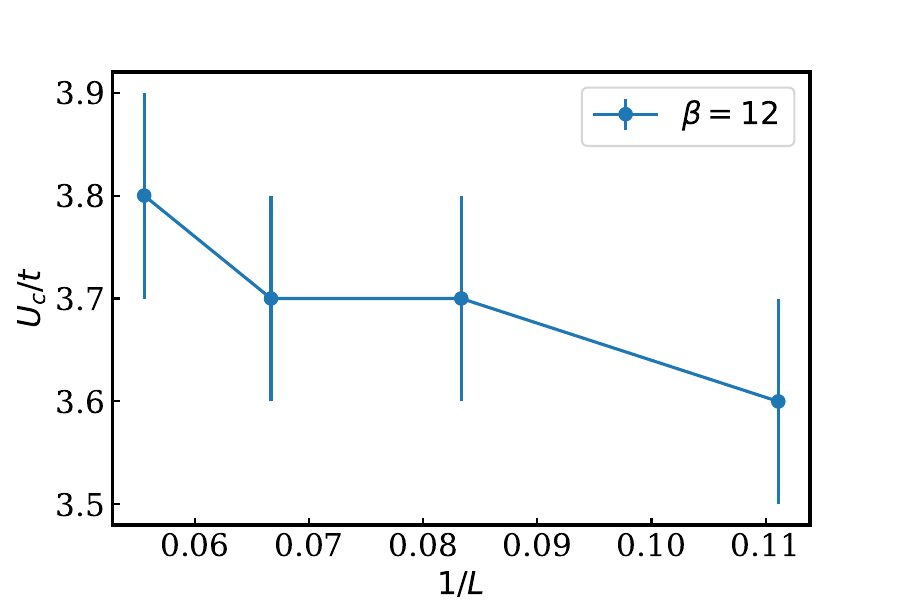}
    \caption{Quantum critical points of the Hubbard model on the honeycomb lattice from Topological Data Analysis, at low temperature and as a function of lattice size.}
    \label{Fig:beta12}
\end{figure}

It is important emphasizing that the result $U_{c}/t \approx 3.7(1)$ obtained from TDA has an relative error $\epsilon =  4(3)\%$, when compared to the FSS case.
This is a reasonable estimation for $U_{c}/t$, considering that we examined much smaller system sizes.
To further examine finite-size dependencies, we performed an additional analysis for $L=18$, at fixed $\beta t =12$, as displayed in Figure \ref{Fig:beta12}.
Notice that, for $L=18$, the QCP predicted by TDA is improved, being $U_{c} = 3.8(1)$, in very good agreement with FSS analysis of Ref.\,\cite{otsuka16}. This suggests that TDA slightly underestimated the QCPs when dealing with small lattice sizes, but the algorithm becomes increasingly more precise for larger values of $L$.
Finally, it is worth comparing TDA with previous results in literature using PCA.
The latter provides a broad crossover for the leading principal component as a function of $U$, with the best estimation for the critical point being $U_{c}/t \approx 4.4$, which is much larger the TDA case.

\section{Conclusions}
\label{sec:concl}

We have implemented a pipeline based on Topological Data Analysis, a machine learning technique coming from Computational Topology.
It extracts powerful information-carrying topological features (persistence diagrams) from point clouds, through which we can distinguish the different phases of a fermionic system.
Within this methodology, we investigated two paradigmatic quantum phase transitions: (i) the AFM-singlet phase transition in the 2D periodic Anderson model, and (ii) the metal-to-insulator one in the Hubbard model on the half-filled honeycomb lattice.
To this end, we performed unbiased determinant quantum Monte Carlo simulations, feeding the TDA algorithm with snapshots of the Hubbard-Stratonovich auxiliary fields, during the course of the simulations.

As our key results, we found a very good agreement with the critical points provided by standard FSS analysis reported in the literature.
In particular to the PAM, TDA have provided the exact same results, within error bars.
Furthermore, we also found that, despite this method exhibits finite size effects, its relative error to the exact critical point is small, which allows one to probe the critical region with reasonable accuracy.
Therefore, we expect that this promising methodology could be a useful tool when dealing with problems that performing a FSS analysis remains a challenge, e.g.~for QMC simulations that exhibit sign problem, or at three-dimensional systems, or even at classical problems that have frustration effects.  

As a final remark, we should mention that the use of machine learning techniques to circumvent, or at least alleviate, the sign problem is an issue of great interest\,\cite{Broecker17}.
However, we have a bottleneck: may our QMC simulations provide useful ``information'' to the machine learning when the sign problem is present? Indeed, this is an open question.
Despite this, we recall that TDA is fed with raw Hubbard-Stratonovich fields, i.e.~it does not involve the computation of average quantities, which inherently requires a reweight with the average sign, $\langle \, \hat{O} S \, \rangle/\langle \, S \, \rangle$.
Therefore, our naive expectation is that further insights into phase transitions could be given even in presence of sign problem, in particular when examining the HS fields with an accurate method, such as TDA.
But, assessing such problems goes beyond the main scope of this paper, and we let these investigations to future works.

\section*{ACKNOWLEDGMENTS}
The authors are grateful to Prof.~S.~Sorella for the discussions. 
A.T.~acknowledges financial support from the MIUR Progetti di Ricerca di Rilevante Interesse Nazionale (PRIN) Bando 2017 - grant 2017BZPKSZ.
N.C.C.~acknowledges the Brazilian Agencies National Council for Scientific and Technological Development (CNPq), National Council for the Improvement of Higher Education (CAPES), and FAPERJ for partially funding this project.
A.T.~and N.C.C.~acknowledge CINECA for awarding them access to the Marconi100 supercomputer, through the ISCRA framework, within the projects AI-H-QMC - HP10BGJH1X, and IsB23 (ISCRA-HP10BF65I0).

\bibliography{ref.bib}

\end{document}